\documentclass[conference,letterpaper]{IEEEtran}

\usepackage[
top    = 0.54in,
bottom = 0.83in,
left   = 0.54in,
right  = 0.54in]{geometry}

\usepackage[utf8]{inputenc} 
\usepackage[T1]{fontenc}
\usepackage{url}
\usepackage{ifthen}
\usepackage{cite}
\usepackage[cmex10]{amsmath} % Use the [cmex10] option to ensure complicance
\usepackage{amssymb}
\usepackage{graphicx}
\usepackage{bm}
\usepackage{bbm}
\usepackage{color}
\usepackage{subfig}
\usepackage{mathtools}

\newtheorem{example}{Example}[section]
\newtheorem{lemma}{Lemma}[section]
\newtheorem{theorem}{Theorem} %[section]

\newtheorem{definition}{Definition}[section]
\newtheorem{remark}{Remark}[section]

\newtheorem{fact}{Fact}[section]

\newcommand{\lp}{\left(}
\newcommand{\rp}{\right)}
\newcommand{\lb}{\left[}
\newcommand{\rb}{\right]}

\newcommand{\mb}{\mathbf}

\newcommand{\Burstfx}{\Burst{n}{\ell}{\thresholds} }
\newcommand{\Burstbnd}{\Burst{n}{\leq\ell}{\thresholds} }
\newcommand{\Burst}[3]{\textsf{$B(#1,#2,#3)$}}
\newcommand{\mini}[1]
{m_{#1}^{*}}
\newcommand{\burst}[1]
{\bm{#1}}
\newcommand{\burstwh}[2]{\bm{#1}^{#2}}

\newcommand{\head}[1]
{h_{\burst{#1}}}
\newcommand{\tail}[1]
{t_{\burst{#1}}}
\newcommand{\length}[1]
{\ell_{\burst{#1}}}
\newcommand{\threshold}[1]
{\eta_{#1}}
\newcommand{\thresholds}{\underline{\eta}}

\newcommand{\height}[1]
{h_{#1}}
\newcommand{\width}[1]
{w_{#1}}
\newcommand{\reverse}[1]{\mathcal{R}\lp #1\rp}
\newcommand{\rep}[2]{#1^{#2}}
\newcommand{\sketch}{\mb{K}}
\newcommand{\refinement}{\mb{R}}
\newcommand{\outcome}{\mb{O}}
\newcommand{\rec}[1]{\mb{M}\lb #1\rb}
\newcommand{\recplm}[1]{\mb{M}^{'}\lb #1\rb}
\newcommand{\recplmTr}[1]{\mb{M}_{r}^{'}\lb #1\rb}
\newcommand{\recplmTl}[1]{\mb{M}_{l}^{'}\lb #1\rb}
\newcommand{\recplmTlr}[1]{\mb{M}_{lr}^{'}\lb #1\rb}
\newcommand{\gray}[2]{\mb{G}_{#1, #2}}
\newcommand{\pgray}[2]{\mb{P}_{#1, #2}}
\newcommand{\graypls}[3]{\mb{\bar{G}}_{#1, #2,#3}}

\newcommand{\B}{\mb{B}}
\newcommand{\shiftleft}[1]{\reflectbox{\ensuremath{\vec{\reflectbox{\ensuremath{#1}}}}}}

\newcommand{\llo}[1]{\reflectbox{\ensuremath{\vec{\reflectbox{\ensuremath{B}}}}}_{\ell}\lp #1\rp}
\newcommand{\oll}[1]{\vec{B}_{\ell}\lp #1\rp}
\begin{document}
%\title{On the Noisy Coin Weighing Problem} 
%\title{On the Noisy Combinatorial Quantitative Group Testing Problem}
%\title{On the Combinatorial Quantitative Group Testing Problem with Noisy Query and Partial Recovery}
\title{Finding a Burst of Positives via Nonadaptive Semiquantitative Group Testing}

\author{
\IEEEauthorblockN{Yun-Han Li$^{a}$, Ryan Gabrys$^{b}$, Jin Sima$^{a}$, Ilan Shomorony$^{a}$ and Olgica Milenkovic$^{a}$
}
\IEEEauthorblockA{
%\IEEEauthorrefmark{1}
$a. $Department of Electrical and Computer Engineering, University of Illinois Urbana-Champaign, USA\\ 
Email: \url{{yunhanl2,jsima,ilans,milenkov}@illinois.edu}
}
\IEEEauthorblockA{
%\IEEEauthorrefmark{1}
$b.$Naval Information Warfare Center,
San Diego\\
Email: \url{ryan.gabrys@gmail.com}
}
}

\maketitle

% abstract

\begin{abstract}
Motivated by testing for pathogenic diseases we consider a new nonadaptive group testing problem for which: (1) positives occur within a burst, capturing the fact that infected test subjects often come in clusters, and (2) that the test outcomes arise from semiquantitative measurements that provide coarse information about the number of positives in any tested group.
% Motivated by group testing applications 
% in genotyping, and in particular, 
% such as COVID-19 testing, where positive cases tend to occur in bursts, we consider the problem of identifying 
% we address the problem of
% identifying a single burst of positive (infected) individuals in a pool of linearly ordered test samples using non-adaptive semiquantitative measurements. 
Our model generalizes prior work on detecting a single burst of positives with classical group testing~\cite{colbourn1999} as well as work on semiquantitative group testing (SQGT)~\cite{emad2014semiquantitative}.
% which has until now not addressed the issue of bursty positives.
Specifically, we study the setting where the burst-length $\ell$ is known and the semiquantitative tests provide potentially nonuniform estimates on the number of positives in a test group. The estimates represent the index of a quantization bin containing the (exact) total number of positives, for arbitrary thresholds $\eta_1,\dots,\eta_s$. Interestingly, we show that the minimum number of tests needed for burst identification is essentially only a function of the largest threshold $\eta_s$. In this context, our main result is an order-optimal test scheme that can recover any burst of length $\ell$ using roughly $\frac{\ell}{2\threshold{s}}+\log_{s+1}(n)$ measurements. 
This suggests that a large saturation level $\eta_s$ is more important than finely quantized information when dealing with bursts. We also provide results for related modeling assumptions and specialized choices of thresholds. 
% Additionally, we provide extensions to the case where only an upper bound on the burst length is known.
% When the burst length is known to be $\ell$, we give an optimal scheme for any SQGT quantized thresholds satisfied $l\geq \threshold{s}\log_{2}(\threshold{s})$ with $\approx\frac{l}{2\threshold{s}}+\log_{s+1}(n)$ tests. Surprisingly, there is no loss to only use the single biggest threshold $\threshold{s}$ when $l=\omega\lp\threshold{s}\log_{2}(n)\rp$. When the upper bound of burst length is known ($\leq l$), we give an optimal scheme for Saturated SQGT quantized thresholds with $\approx\frac{l}{\threshold{s}}+\log_{2}(n)$ tests.
\end{abstract}

%\noindent\textit{An extended version of this paper is accessible at:}\\
%\centerline{\url{http://homepage.ntu.edu.tw/~ihwang/Eprint/itw20cqgt.pdf}}

\section{Introduction}\label{sec:intro}
Group testing (GT) is a protocol for identifying relatively small subsets of marked elements, referred to as \emph{positives}, within a larger collection of entities termed test subjects. The gist of the approach is to group subjects into carefully selected subpools and test the subjects in each subpool jointly so as to reduce the number of tests compared to that needed for individual testing. The first GT scheme comprising two stages of testing was described by Dorfman~\cite{dorfman} in the context of finding individuals with venereal diseases. His scheme also represents the first instance of \emph{adaptive testing}, where measurements from one round of testing can be used to inform the test selections in subsequent rounds. Unlike adaptive testing, \emph{nonadaptive} GT requires that all tests be designed and conducted simultaneously. Since Dorfman's work, GT has been extended and generalized in many different directions and has found numerous applications in search systems, experimental and circuit design and computational biology. For comprehensive surveys, the interested reader is referred to~\cite{GT-ITperspective},\cite{combinatorial2000}.

In~\cite{colbourn1999}, Colbourn considered a specialized GT technique for identifying one single burst of consecutive positives of length $\leq \ell$ within an ordered list of $n$ elements. For nonadaptive techniques, Colbourn showed that the order-optimal number of measurements equals $\ell+\log(n)$. Follow-up works focused on improving some aspects of the scheme~\cite{lin2013synthetic,muller2004consecutive,2021improved}, extending the results to include new adaptive protocols~\cite{juan2008adaptive}, and generalizing the approach to handle multiple bursts~\cite{mutipleburst}.

However, in many real-life scenarios, such as testing for infections with viral pathogens based on quantitative PCR (quantitative polymerase chain reaction, qPCR), the outcomes are real-valued and usually confined to an interval such as $[10,45]$. A measurement is known as the $C_t$ (cycle threshold) value and it conveys information about how likely an individual is to be infected. For example, a $C_t$ value close to $40$ is highly indicative of a negative subject, while a value below $20$ is a strong sign that the individual is highly virulent. One can therefore quantize the $C_t$ values using a carefully selected collection of $s$ thresholds $\underline{\eta}=(\eta_1,\ldots,\eta_s)$ so that each quantization bin provides an estimate of the viral load in the pool and, consequently, an estimate of the number of positives in the pool. This type of GT approach is known as semiquantitative GT (SQGT)~\cite{emad2014semiquantitative,emad2016code}. Furthermore, whenever testing is done on large populations in which individuals that cohabitate are naturally adjacent in the order used for testing~\cite{lin2021positively} (for example, families, dorm-mates etc.), bursty positive models are appropriate and can result in significant savings compared to classical GT approaches~\cite{colbourn1999}.

Given the additional quantitative information and the assumption regarding consecutive orderings of positives, one can easily envision performing SQGT for bursty positive that quantizes the $C_t$ values into quantization bins that indicate the level of the viral load, or an estimate of the number of infected individuals in the population~\cite{AC-DC,cheraghchi2021semiquantitative}. Here, for the first time, we study the reduction in the number of group measurements achievable in such a setting. In particular, we investigate two new bursty SQGT models~\cite{emad2014semiquantitative}, one in which the length of the burst is known and fixed to $\ell$ (henceforth referred to as the fixed-length burst model, $\Burstfx$); and another, in which the length of the burst is known to be upper-bounded by $\ell$ (henceforth referred to as the bounded-length burst model, $\Burstbnd$).
    
Our main contributions include 
    \begin{enumerate}
        \item Order-optimal constructions (i.e., constructive lower and upper bounds that differ by a constant factor of $2$) for the $\Burstfx$ setting with quantization thresholds $\thresholds$ for which $\ell=\Omega\lp\threshold{s}\log_{2}(\threshold{s})\rp$.
        \item Order-optimal constructions (i.e., lower and upper bounds that differ by a constant factor of $4$) for the $\Burstbnd$ setting with SQGT thresholds $\thresholds=(1,\hdots,s)$ corresponding to the so-called \emph{saturation model}~\cite{dyachkov1983survey,AC-DC}.
    \end{enumerate}
Two important comments are in place. Semiquantitative measurements significantly decrease the number of tests needed for the $\Burstbnd$ setting (the improvement is linear in the number of thresholds $s$). Somewhat surprisingly, for the $\Burstfx$ setting the number of tests is basically determined by the value of the largest threshold $\threshold{s}$ rather than by the total number of thresholds $s$. These findings may have interesting consequences for test schedules and quantization schemes used for practical quantitative PCR protocols. 

The paper is organized as follows. Section II introduces the notation and provides the formal problem formulation. Section III contains the results for the lower bounds, while Section IV contains the main results of the work, pertaining to upper bounds on the number of SQGT burst identification models for a fixed and upper-bounded length of the burst. %Remarks and open problems are presented in Section V. 

\section{Problem Formulation}\label{sec:formulation}

We start by introducing the relevant notation as well as the fixed-length and bounded-length single burst identification problems under SQGT measurements.

Let $\height{\mb{M}}, \width{\mb{M}}, \mb{M}\lp*,i\rp, \mb{M}\lp j,*\rp$ denote the number of rows (height), number of columns (width), $i$-th column and $j$-th row of the matrix $\mb{M}^{\height{\mb{M}}\times\width{\mb{M}}}$, respectively. Our row indices lie in $\lb0,\height{\mb{M}}-1\rb$, while the column indices are confined to $\lb0,\height{\mb{M}}-1\rb$. In addition, $\reverse{\mb{M}}, \rep{\mb{M}}{c}, \rep{\mb{M}}{\infty}$ are used to denote a matrix obtained from $\mb{M}$ by reversing the column order (so that $\reverse{\mb{M}}(*,i)=\mb{M}(*,\width{\mb{M}}-i-1)$, a $c$-fold horizontal concatenation of $\mb{M}$ (i.e, $\lb\mb{M},\hdots,\mb{M}\rb$ with $c$ constituent matrices), and a horizontal concatenation of matrices $\mb{M}$ such that $\width{\mb{M}{\infty}}$ becomes a value specified during the construction process, respectively. Finally, we use $\mb{M}(i,j)$ to denote the entry in $\mb{M}$ in row $i$ and column $j$. 
%$\bm{v}_{j}$ denote the $j$-th element of a vector $\bm{v}$. 

The single burst of positives problem requires introducing the following notions.

\underline{Bursts}: A burst is denoted by a binary $n\times 1$ column vector $\burst{b}$, and is specified by a head and tail $\head{b}\leq \tail{b}$, which dictate its length $\length{b}=\tail{b}-\head{b}+1$. It comprises consecutive positives:
\begin{equation*}
   \burst{b}(i)=
    \begin{cases}
        0,& 0\leq i<\head{b},\\
        1,& \head{b}\leq i\leq\tail{b},\\
        0,& \tail{b}<i\leq n.
    \end{cases}   
\end{equation*}      
When $\length{b}$ is fixed, $\burstwh{b}{i}$ denotes the burst with $\head{b}=i$, and the distance between two burst $\burstwh{b}{i},\burstwh{b}{j}$ is defined as the difference of their head position $|i-j|$.
\par%\smallskip\noindent
\underline{SQGT measurements}: 
An SQGT measurement is described by 
a $1\times n$ binary vector $\bm{m}$ such that
\begin{equation*}
    \bm{m}\lp i\rp=
    \begin{cases}
        1,& i\text{th element is included in the test},\\
        0,& \text{otherwise},        
    \end{cases}        
\end{equation*}
and a set of integer-valued quantized thresholds
\begin{equation*}
    \thresholds=\lp\threshold{1},\hdots,\threshold{s}\rp\text{ with }0<\threshold{1}<\hdots<\threshold{s}\leq n,
\end{equation*}
such that the SQGT measurement outcomes equal
\begin{equation*}
    \thresholds(\bm{m}\burst{b})=
    \begin{cases}
        0,& 0\leq\bm{m}\burst{b}<\threshold{1},\\
        i,& \threshold{i}\leq\bm{m}\burst{b}<\threshold{i+1},\\
        s,& \threshold{s}\leq\bm{m}\burst{b}\leq n.
    \end{cases}
\end{equation*}
\begin{definition}
    When $\thresholds=(1,2,\hdots,s)$, we refer to this specialized SQGT scheme as the \emph{saturation} SQGT model. 
\end{definition} 
\par%\smallskip\noindent
\underline{Correct burst detection}: for any hidden burst $\burst{b}$, the estimate generated by the detection algorithm, denoted by $\hat{\burst{b}}$, should  equal $\burst{b}$. 
\par%\smallskip

%We consider two different single-burst problems in this paper. 
\begin{definition}
   \Burstfx and \Burstbnd are used to denote the fixed-length 
    and bounded-length burst problem with burst-lengths $=\ell$ and $\leq\ell$, respectively, and with $n$ test elements and SQGT quantized thresholds $\thresholds$. 
\end{definition}

A nonadaptive SQGT testing scheme with $m$ measurements on $n$ elements is represented by a $m\times n$ binary measurement matrix $\mb{M}$ with each row corresponding to a single SQGT measurement.
We say $\mb{M}$ solves the \Burstfx (or the \Burstbnd) problem if and only if
\begin{align*}
    \forall \burst{b}\neq\burst{b^{'}}& \text{ allowed by the \Burstfx (\Burstbnd) problem},\\ & \text{  one has } \thresholds(\mb{M}\burst{b})\neq \thresholds(\mb{M}\burst{b^{'}}).
\end{align*}
The smallest possible number of measurements possible to meet this requirement, among all nonadaptive SQGT schemes is 
denoted by $\mini{\Burstfx}$ and $\mini{\Burstbnd}$.

Our constructions will make use of Gray codes and generalizations thereof. We say that
$\gray{s}{h}\in\{0,\hdots,s\}^{h\times s^{h}}$ represents an $s$-ary Gray code with length $h$ if it satisfies the following two conditions:
\begin{enumerate}
    \item Any two consecutive columns differ in exactly one position, and the difference has magnitude one.
    \item $\gray{s}{h}$ includes all possible $s^{h}$ codewords exactly once.
\end{enumerate}
\begin{example}\label{gray_example}
The following matrix has columns that constitute a $3$-ary Gray code of length two:
\begin{align*}
\begin{bmatrix}
0 & 0 & 0 & 1 & 1 & 1 & 2 & 2 & 2 \\
0 & 1 & 2 & 2 & 1 & 0 & 0 & 1 & 2
\end{bmatrix}.
\end{align*}
\end{example}

\begin{fact}
    The Gray code $\gray{s}{h}$ can be constructed by first recursively constructing paired Gray code matrices $\pgray{s}{h}:=\lb\gray{s}{h},\reverse{\gray{s}{h}}\rb$ using the rule below and then removing half of the columns from the right side:
    \begin{equation}\label{graycode_rule}
        \begin{cases}
            \pgray{s}{1}=\lb 0,\hdots,s-1,s-1,\hdots,0\rb,\\
            \pgray{s}{i}=\begin{bmatrix}
                \pgray{s}{1}\otimes \bm{1}^{s^{i-1}}\\
                \rep{\pgray{s}{i-1}}{s}.
            \end{bmatrix}
        \end{cases}
    \end{equation}
\end{fact}
Here, $\otimes$ stands for the Kronecker product while $\bm{1}^a$ is a row vector of $1$s.
\begin{example}
    The following matrix $\pgray{3}{2}$ is constructed recursively using (\ref{graycode_rule}). The left half, as claimed, equals $\gray{3}{2}$ and was  illustrated in Example~\ref{gray_example}:
    \begin{align*}
    \begin{bmatrix}
    000 & 111 & 222 & 222 & 111 & 000\\
    012 & 210 & 012 & 210 & 012 & 210
    \end{bmatrix}.
    \end{align*}
\end{example}
%\textcolor{blue}{\bf do we need to change the second half of the lower row above to be 210 012 210?}
We also make use of the following property of binary Gray codes.
\begin{fact}\label{gray_num_burst}
    $\gray{2}{h}\lp(i,*)\rp$ contains $2^{i-1}$ runs of $1$s for all $i$ except $i=0$, which contains only one run of $1$s. Consequently, the matrix contains a total of $\sum_{i=1}^{h-1}2^{i-1} + 1=2^{h-1}$ runs of $1$s within its rows. This is illustrated by the following example for $\gray{2}{3}$, with a total number of $2^{3-1}=4$ runs of $1$s.
    \begin{align*}
    \begin{bmatrix}
    0 & 0 & 0 & 0 & [1 & 1 & 1 & 1]\\
    0 & 0 & [1 & 1 & 1 & 1] & 0 & 0\\
    0 & [1 & 1] & 0 & 0 & [1 & 1] & 0
    \end{bmatrix}.
    \end{align*}
\end{fact}

\section{Lower Bounds}\label{sec:limits}
%\subsection{Fundamental limits for the non-adaptive case}
We first provide lower bounds for the smallest number of measurements needed for the $\mini{\Burstfx}$ and $\mini{\Burstbnd}$ settings. The proofs mostly use ideas from~\cite{colbourn1999}.
\begin{theorem}\label{LB_fix}\label{LB_variable} We have
    \begin{equation*}
        \begin{cases}
        \mini{\Burstfx}\geq \max\lp\log_{s+1}\lp n-\ell+1\rp,\lceil\frac{\ell}{2\threshold{s}}\rceil\rp,\\
        \mini{\Burstbnd}\geq \max\lp\log_{2}\lp n\rp,\lceil\frac{\ell}{\threshold{s}}\rceil\rp.
        \end{cases}   
    \end{equation*}
\end{theorem}

The proof technique used for $\mini{\Burstbnd}$ is similar to that for $\mini{\Burstfx}$; hence, we only provide the proof for $\mini{\Burstfx}$.   
We prove the first bound by establishing each of the bounds on the right-hand side separately and combining them via maximization. 
\begin{enumerate}
    \item The bound $\log_{s+1}\lp n-\ell+1\rp$ follows from a simple counting argument: there are a total of $n-\ell+1$ different head positions and a total of $s+1$ possible outcomes for each measurement.
    \item The bound $\lceil\frac{\ell}{2\threshold{s}}\rceil$: we show that even if we only require to discriminate among the first $\ell+1$ bursts (i.e., bursts $\burstwh{b}{i}$ with $0\leq i\leq\ell$), we still need $\lceil\frac{\ell}{2\threshold{s}}\rceil$ measurements. For any measurement $\bm{m}$, let $\bm{m}_{1}$ and $\bm{m}_{2}$ denote the first and second block of $\ell$ bits of $\bm{m}$. Only the last $\threshold{s}$ nonzero bits in $\bm{m}_{1}$ and the first $\threshold{s}$ nonzero bits in $\bm{m}_{2}$ are relevant. For simplicity, we only provide a proof for the $\bm{m}_{2}$ case. Let $\ell\leq i_{1}<i_{2}<\hdots<2\ell$ be the elements included in $\bm{m}_{2}$. Since $\length{b}=\ell$ and $\head{b}\leq\ell$, if $i_{j}$ is included in the burst $\burst{b}$ then $i_{1},\hdots,i_{j-1}$ must also be included. Therefore, if $j>\threshold{s}$, by observing that $\threshold{s}$ is the largest threshold, one can remove $i_{j}$ from $\bm{m}_{2}$ without changing the outcome. Hence one can only retain the first $\threshold{s}$ nonzero bits in $\bm{m}_{2}$ and still arrive at the same outcome. As a result, it suffices to only consider those $\bm{m}$ for which $\sum_{j=0}^{2\ell-1}\bm{m}(j)\leq 2\threshold{s}$. Let $\mb{M}^{h_{\mb{M}}\times 2\ell}$ be our measurement matrix restricted to the first $2\ell$ columns. Suppose that $h_{\mb{M}}<\frac{\ell}{2\threshold{s}}$; then $\sum_{i=0}^{h_{\mb{M}}-1}\sum_{j=0}^{2\ell-1} \mb{M}(i,j)\leq 2\threshold{s}h_{\mb{M}}<\ell$. This implies that there exists a $0\leq j<\ell$ such that $\mb{M}(*,j)=\mb{M}(*,j+\ell)=\bm{0}^{\height{\mb{M}}\times 1}$. Then $\thresholds\lp\mb{M}\burstwh{b}{i+1}\rp= \thresholds\lp\mb{M}\burstwh{b}{i}+\mb{M}(*,j+\ell)-\mb{M}(*,j)\rp=\thresholds\lp\mb{M}\burstwh{b}{i}\rp$. Therefore $\mini{\Burstfx}\geq \lceil\frac{\ell}{2\threshold{s}}\rceil$. 
\end{enumerate}

\section{Main Results}\label{sec:results}
In Section~\ref{General_SQGT_fix}, we describe and order-optimal construction of measurement matrices for the \Burstfx problem pertaining to two different cases, the case of general SQGT thresholds with $\ell=\Omega(\threshold{s}\log(\threshold{s}))$ and the saturation model with $\ell\leq\threshold{s}=s$. It is interesting to note that for the first case, $\mini{\Burstfx}$ basically depends only on the largest threshold $\threshold{s}$. In other words, as long as $\ell=\Omega(\threshold{s}\log_{2}(n))$ with sufficiently large constant, there is no benefit of using multiple thresholds (SQGT) compared to threshold group testing (TGT) with the single biggest threshold $\threshold{s}$. In Section~\ref{Saturated_SQGT_bounded}, we describe an order-optimal scheme (within an approximation constant $4$) for \Burstbnd problem and the saturation model.
%Then, we propose a construction that can achieve the optimal leading coefficient when efficient algorithms for \SCQGT{n,n^{\kappa},n^{\delta},n^{\lambda}} are available. 
%To this end, the design problem boils down to finding good measurement plans for \SCQGT{n,n^{\kappa},n^{\delta},n^{\lambda}}. We give the explicit construction of a sufficiently good adaptive measurement plan for \SCQGT{n,n^{\kappa},n^{\delta},n^{\lambda}} in Section~\ref{adaptive_embedding}. While the explicit construction of an efficient non-adaptive measurement plan for \SCQGT{n,n^{\kappa},n^{\delta},n^{\lambda}} is still missing, we provide reduction of this problem to a simpler combinatorial problem detailed in Section~\ref{reduction_SCQGT}.   
\vspace{-0.05in}

\subsection{The \Burstfx Model}\label{General_SQGT_fix}

Since for this case the burst length is fixed, one only needs to locate the position of the head $\head{b}\in[0,n-\ell]$ of the burst $\burst{b}$.
Vaguely speaking, the near-optimal construction follows a two-step sketch-and-refine procedure. The first part, referred to as the \emph{General Sketch Scheme}, uses a measurement matrix $\sketch$ (Theorem~\ref{General_Sketch}) that distinguishes bursts separated by $>\ell+1$ positions. The second part, referred to as the \emph{General Refinement Scheme}, uses a measurement matrix $\refinement$ (Theorem~\ref{General_Refined}) that distinguishes bursts separated by $<2\ell$ positions. Stacking the two measurement matrices leads to the result reported in Theorem~\ref{Optimal_SQGT_scheme}.

\begin{theorem}\label{General_Sketch}
For \Burstfx, the measurement matrix $\sketch$ described in Section~\ref{proof_General_Sketch} can distinguish all bursts at distances $>\ell+1$ using $\lceil\log_{s+1}\lp \frac{n-\ell+1}{\ell}\rp\rceil$ measurements.
\end{theorem}

\begin{theorem}\label{General_Refined}
    For \Burstfx with parameters $\threshold{s}=2^{h-1}+2$ and $\ell=c2^{h}+1$, where $c,h\in\mathbb{N}$ and $c>2(h+1)$, the measurement matrix $\refinement$ described in Section~\ref{proof_General_refinement} can distinguish all bursts at distances $<2\ell$ using roughly 
    $\frac{\ell}{2\threshold{s}}$ measurements. The scheme depends only on the largest threshold $\threshold{s}$.   
\end{theorem}

\begin{theorem}\label{Optimal_SQGT_scheme}
 Combining the General Sketch matrix of Theorem~\ref{General_Sketch} and the General Refinement matrix of Theorem~\ref{General_Refined}, leads to the measurement matrix $\lb\mb{K}^{\intercal},\mb{R}^{\intercal}\rb^{\intercal}$
which can be used to solve \Burstfx using roughly $\frac{\ell}{2\threshold{s}}+\log_{s+1}\lp \frac{n-\ell+1}{\ell}\rp$ measurements. This number of measurement is at most twice the number of measurement from the lower bounds reported in Theorem~\ref{LB_fix}.
\end{theorem}

\begin{remark}\label{remark_interesting}
Note that scheme from Theorem~\ref{General_Refined} only uses the largest threshold $\threshold{s}$. Therefore, if we only make use of  $\threshold{s}$ in the General Sketch Scheme of Theorem\ref{General_Sketch}, the resulting measurement matrix has height roughly $ \frac{\ell}{2\threshold{s}}+\log_{2}\lp \frac{n-\ell+1}{\ell}\rp$ and depends on one threshold, $\threshold{s}$. When $\ell=\Omega(\threshold{s}\log_{2}(n))$ with sufficiently large constant, $\frac{\ell}{2\threshold{s}}+\log_{s+1}\lp \frac{n-\ell+1}{\ell}\rp=\Omega\lp\log_{2}(n)\rp=\frac{\ell}{2\threshold{s}}+\log_{2}\lp \frac{n-\ell+1}{\ell}\rp$. Therefore, in this parameter regime, there is no benefit from using multiple thresholds.
\end{remark}

\subsubsection{Proof of Theorem~\ref{General_Sketch}\label{proof_General_Sketch}} We start with some relevant notation. Let $\sketch^{h_{\sketch}\times n}$ be the measurement matrix. We say that $\sketch^{h_{\sketch}\times n}$ results in the outcome matrix $\outcome^{h_{\sketch}\times n-\ell+1}$ if $\outcome=\lb\thresholds\lp\sketch\burstwh{b}{0}\rp \hdots \thresholds\lp\sketch\burstwh{b}{n-\ell+1}\rp\rb$ represents the collection of outcomes for all length-$\ell$ bursts $\burstwh{b}{i}$ when using the measurement matrix $\sketch$. 

Next, let $\oll{x}:=\bm{0}^{\ell-x}\bm{1}^{x}$ and $\llo{x}:=\bm{1}^{x}\bm{0}^{\ell-x},$ for all $x\in\{0,\hdots,\ell\}$. Also, let $\rep{\oll{0}}{i},\rep{\llo{0}}{i}$ stand for the horizontal concatenation of $i$ copies of $\oll{x}$ and $\llo{x}$. Observe that $\lceil\log_{s+1}\lp \frac{n-\ell+1}{l}\rp\rceil$ (almost) matches the counting bound $\log_{s+1}\lp \frac{n-\ell+1}{\ell}\rp$. The key idea is to first construct $\sketch$ with $\width{\sketch}\geq n$ such that the outcome matrix satisfies
\vspace{-3mm}
\begin{equation}
    \outcome= \gray{s+1}{\height{\sketch}}\otimes\mb{1}^{\ell+1},\vspace{-3mm}
\end{equation}
and then truncate it to $n$ columns. By the definition of $\outcome$, $\sketch$ can identify all bursts at distance $\geq\ell+1$ if and only if all columns of $\gray{s+1}{\height{\sketch}}$ are different; that this is true follows from the fact that Gray codes include all vectors $\{0,\hdots,s\}^{\height{\sketch}}$ exactly once. We need the following lemma for our subsequent derivations.
\begin{lemma}\label{repeat_lemma}
The following measurement%\vspace{-3mm}
    \begin{align*}
        &\bm{m}\lp c\rp:=\left[\rep{\oll{0}}{c}  0\rep{\oll{\threshold{1}}}{c}\hdots  0\rep{\oll{\threshold{s}}}{c}\right. \\
        &\left. 
        1\rep{\llo{\ell}}{c}
        1\rep{\llo{\threshold{s}-1}}{c} \hdots 1\rep{\llo{\threshold{1}-1}}{c} 0\right]^{\infty}
    \end{align*}
    results in the outcome
    $\rep{\lp\lb 0,\hdots,s,s,\hdots,0\rb\otimes \bm{1}^{c\ell+1}\rp}{\infty}$.
\end{lemma}
\begin{IEEEproof} The case $c=1$ can be proved easily and is illustrated by the following example. For $\thresholds=(1,2,4)$ 
and $\ell=6$, $\bm{m}(1)$ equals       
\begin{align*}
(000000&\ 0000001\ 0000011\ 0001111\\
1111111&\ 1111000\ 1100000\ 1000000\ 0)
\end{align*}   
For $c>1$, and any length-$\ell$ row-vector $\bm{x}$, $\thresholds\lp\rep{\bm{x}}{c}\burstwh{b}{i}\rp$ remains unchanged for all $\burstwh{b}{i}$, where $i \in [0,(c-1)\,\ell]$.
    \begin{equation*}
        \forall i, \; \thresholds\lp\rep{\bm{x}}{c}\burstwh{b}{i+1}\rp = \thresholds\lp\rep{\bm{x}}{c}\burstwh{b}{i}+\rep{\bm{x}}{c}(*,i+\ell)-\rep{\bm{x}}{c}(*,i)\rp = \thresholds\lp\rep{\bm{x}}{c}\burstwh{b}{i}\rp
    \vspace{-2mm}
    \end{equation*} 
    \vspace{-1mm}
\end{IEEEproof}
We are now ready to present our construction. Let $\rec{i}$ be the measurement matrix recursively constructed as follows:
\begin{equation}\label{rec_rule}
    \begin{cases}
        \rec{1}=\bm{m}\lp 1\rp,\\
        \rec{i}=\begin{bmatrix}
            \bm{m}\lp \frac{(\ell+1)(s+1)^{i-1}-1}{\ell}\rp\\
            \rep{\rec{i-1}}{s+1}
        \end{bmatrix}.
    \end{cases}
\end{equation}
Note that $\frac{(\ell+1)(s+1)^{i-1}-1}{\ell}$ may not be an integer. We therefore first focus on the special case $s=\ell$ (therefore $\frac{(\ell+1)(s+1)^{i-1}-1}{\ell}=\frac{(\ell+1)^{i}-1}{\ell}\in\mathbb{N}$) and then generalize the result for $s<\ell$ through slight modifications of the argument.
\begin{lemma}\label{rec_lemma}
    For $s=\ell$, $\rep{\rec{i}}{\infty}$ results in the outcome matrix $\rep{\lp\pgray{\ell+1}{i}\otimes \bm{1}^{\ell+1}\rp}{\infty}$. Where $\pgray{\ell+1}{i}$ is the  $\ell+1$-ary length-$i$ paired gray code matrix.
\end{lemma}
\begin{IEEEproof} The proof is by induction. 
    \begin{enumerate}
        \item For $i=1$: by Lemma~\ref{repeat_lemma}, $\rep{\rec{i}}{\infty}=\rep{\bm{m}\lp1\rp}{\infty}$ results in the outcome matrix 
        $\rep{\lp\lb 0,\hdots,\ell,\ell,\hdots,0\rb\otimes \bm{1}^{\ell+1}\rp}{\infty}=\rep{\lp\pgray{\ell+1}{1}\otimes \bm{1}^{\ell+1}\rp}{\infty}$.
        \item For $i>1$: Suppose that the claim holds for $i-1$. Then
        \begin{equation*}
            \rep{\rec{i}}{\infty}=\rep{\begin{bmatrix}
                \bm{m}\lp \frac{(\ell+1)^{i}-1}{\ell}\rp\\
                \rep{\rec{i-1}}{\ell+1}
            \end{bmatrix}}{\infty}
        \end{equation*}
        results in the outcome matrix
        \begin{align*}
            &\rep{\begin{bmatrix}
                \lb 0,\hdots,\ell,\ell,\hdots,0\rb\otimes \bm{1}^{(\ell+1)^{i}}\\
                \pgray{\ell+1}{i-1}^{\ell+1}\otimes \bm{1}^{\ell+1}
            \end{bmatrix}}{\infty}=\rep{\lp\pgray{\ell+1}{i}\otimes \bm{1}^{\ell+1}\rp}{\infty}.
        \end{align*}
    \end{enumerate}
\end{IEEEproof}
By Lemma~\ref{rec_lemma}, $\rec{\height{\sketch}}$ truncated to $(\ell+1)^{\height{\sketch}+1}+\ell-1$ columns from the right results in the outcome matrix $\gray{\ell+1}{\height{\sketch}}\otimes \bm{1}^{\ell+1}$, and can consequently 
distinguish all bursts within distance $>\ell+1$. It is not hard to show that a single measurement $\bm{0}^{\ell}\rep{\lp \bm{1}^{\ell+1}\bm{0}^{\ell+1}\rp}{\infty}$ results in the outcome matrix $\rep{\lp0,\hdots,\ell,\ell,\hdots,0\rp}{\infty}$. Hence we have the following theorem.
\begin{theorem}
    For the saturation SQGT model with $\ell$ thresholds $\thresholds=(1,\hdots,\ell)$, 
    $\begin{bmatrix}
        \rep{\rec{\lceil\log_{\ell+1}\lp n-\ell+1\rp-1\rceil}}{\infty}\\
        \bm{0}^{\ell}\rep{\lp \bm{1}^{\ell+1}\bm{0}^{\ell+1}\rp}{\infty}
    \end{bmatrix}$ 
truncated to $n$ columns on the right can be used as the test matrix for the \Burstfx model with $\lceil\log_{\ell+1}\lp n-\ell+1\rp\rceil$ measurements.
\end{theorem}
For the case $s<\ell$, some modifications in the recursion given by (\ref{rec_rule}) are required. The modification involves truncating 
a certain number of columns from the left, right, or both sides of $\recplm{i}$ at each stage of recursion $i$:
\begin{align*}
    \begin{cases}
        \recplm{1}=\bm{m}\lp 1\rp,\\
        \recplm{i}=\begin{bmatrix}
            \bm{0}^{\alpha\bmod\ell}& \bm{m}\lp\lfloor\frac{\alpha}{\ell}\rfloor\rp& \bm{0}^{\alpha\bmod\ell}\\
            \recplmTr{i-1}& \rep{\recplmTlr{i-1}}{s-1}& \recplmTl{i-1}
        \end{bmatrix},
    \end{cases}
\end{align*}
where $\alpha=\frac{\width{\recplm{i-1}}}{2}-1$, and $\recplmTr{i-1},\recplmTl{i-1},\recplmTlr{i-1}$ denotes $\recplm{i-1}$ truncate $\alpha\bmod\ell$ columns from the right, left, and both sides, respectively.

By using a similar proof as the one described above and some simple but tedious calculations, one can show that $\recplm{\lceil\log_{s+1}\lp\frac{n-\ell+1}{\ell}\rp\rceil}$ truncated to $n$ columns from the right can be used as $\sketch$. Hence, $\sketch$ can distinguish all bursts at distance $>\ell+1$ using $\lceil\log_{s+1}\lp \frac{n-\ell+1}{\ell}\rp\rceil$ measurements.

%$\vec{0}$ and $\vec{1}$ denote the all $0$ and all $1$ column vector. We use
\subsubsection{Proof of Theorem~\ref{General_Refined}}\label{proof_General_refinement}
We now focus our attention on the General Refinement Scheme. Let $\shiftleft{\B}$ to denote the cyclic shift of columns in $\B$ one position to the left so that $\shiftleft{\B}(*,i)=\B(*,i+1\bmod\ell)$. We need the following lemma.
\begin{lemma}\label{lemma}
    Suppose that a binary matrix $\B^{\height{\B}\times\ell}$ satisfies the following three conditions:
    \begin{enumerate}
        \item\label{condition1} All columns and their binary complement $\{\B(*,i), \bar{\B}(*,i)\}_{i=0}^{\ell-1}$ are distinct.
        \item\label{condition2} The first column is the zero vector, $\B(*,0)=\bm{0}^{\height{\B}\times1}$.
        \item\label{condition3} Each row of $\shiftleft{\B}-\B$ has $\threshold{s}-1$ elements equal to $-1$.
    \end{enumerate}
    Then the following measurement matrix can be used to distinguish all bursts $\burst{b}\neq\burst{b^{'}}$ at distance $<2\ell$:
    \vspace{-1mm}
    \begin{equation}
        \refinement:=\lb\refinement^{-} \refinement^{+} \refinement^{-} \refinement^{+}\hdots\rb,
        \vspace{-1mm}
    \end{equation}
where $\refinement^{-}$ denotes the ``negative'' part of $\shiftleft{\B}-\B$ (obtained by setting $1$s to $0$s and $-1$s to $1$s), while $\refinement^{+}$ denotes the positive part of $\shiftleft{\B}-\B$ (obtained by setting $-1$s to $0$s), and the last column is changes from $\bm{0}^{\height{\B}\times1}$ to $\bm{1}^{\height{\B}\times1}$ (note that the last column of $\shiftleft{\B}-\B$ before the modification is $\B(*,0)-\B(*,\ell-1)=-\B(*,\ell-1)$, which implies that the positive part is zero).
\end{lemma}
\begin{IEEEproof}
    Since $\refinement$ is a repeated horizontal concatenation of $\refinement^{-}$ and $\refinement^{+}$, it suffices to show that
    \vspace{-1mm}
    \begin{equation}\label{criteria_refinement}
        \forall 0\leq i\neq j<2\ell, \; \thresholds\lp\refinement\burstwh{b}{i}\rp\neq\thresholds\lp\refinement\burstwh{b}{j}\rp.
        \vspace{-1mm}
    \end{equation}
    In particular, we prove that
    \vspace{-1mm}
    \begin{equation}\label{run}
        \refinement\burstwh{b}{i}=\begin{cases}
            \lp\threshold{s}-1\rp\bm{1}^{\height{\B}\times1}+\B(*,i) &0\leq i<\ell,\\
            \threshold{s}\bm{1}^{\height{\B}\times1}-\B(*,i) &\ell\leq i<2\ell.\\
        \end{cases}
        \vspace{-1mm}
    \end{equation}
%\vspace{-5mm}
Note that each entry of $\refinement\burstwh{b}{i}$ is either $\threshold{s}-1$ or $\threshold{s}$. By condition~\ref{condition1}, all $\refinement\burstwh{b}{i}$ are different. Consequently, all $\thresholds\lp\refinement\burstwh{b}{i}\rp$ are different as well. Therefore, $\refinement$ can distinguishes all bursts at distance $<2\ell$ using $\height{\B}$ measurements; only the largest threshold $\threshold{s}$ is relevant. Next we prove~(\ref{run}).

For $0\leq i<\ell$,
\vspace{-3mm}
\begin{align*}
    \refinement\burstwh{b}{i}-\refinement\burstwh{b}{0}
    &=\sum_{j=1}^{i}\refinement\burstwh{b}{j}-\refinement\burstwh{b}{j-1}=\sum_{j=1}^{i}\refinement\lp\burstwh{b}{j}-\burstwh{b}{j-1}\rp\\ 
    &=\sum_{j=1}^{i}\refinement^{+}(*,j-1)-\refinement^{-}(*,j-1)\\
    &=\sum_{j=1}^{i}\refinement(*,j-1)\overset{(1)}{=}\B(*,i).
    \vspace{-5mm}
\end{align*}
    For $\ell\leq i<2\ell$,
    \vspace{-4mm}
    \begin{align*}
        \refinement\burstwh{b}{i}-\refinement\burstwh{b}{\ell}
        &=\sum_{j=\ell+1}^{i}\refinement\burstwh{b}{j}-\refinement\burstwh{b}{j-1}=\sum_{j=\ell+1}^{i}\refinement\lp\burstwh{b}{j}-\burstwh{b}{j-1}\rp\\ 
        &=\sum_{j=\ell+1}^{i}\refinement^{-}(*,j-1)-\refinement^{+}(*,j-1)\\
        &=-\sum_{j=1}^{i-\ell}\refinement(*,j-1)\overset{(1)}{=}-\B(*,i).
        \vspace{-4mm}
    \end{align*} 
The equalities $\overset{(1)}{=}$ follows from Condition~\ref{condition2}. Finally, by Condition~\ref{condition3} and the fact that we changed the last column of $\refinement^{+}$ from $\bm{0}^{\height{\B}\times1}$ to $\bm{1}^{\height{\B}\times1}$, $\refinement\burstwh{b}{0}= \lp\threshold{s}-1\rp\bm{1}^{\height{\B}\times1}$ and $\refinement\burstwh{b}{\ell}-\refinement\burstwh{b}{0}=\bm{1}^{\height{\B}\times1}\Rightarrow \refinement\burstwh{b}{\ell}=\threshold{s}\bm{1}^{\height{\B}\times1}$.
\end{IEEEproof}
It remains to construct a matrix $\B$ that satisfies Conditions~\ref{condition1}-\ref{condition3} in Lemma~\ref{lemma} with $\height{\B}$ roughly equal to $\frac{\ell}{2\threshold{s}}$. Let $\gray{2}{h}^{h\times2^{h}}$ be the code matrix of a binary Gray code of length $h$ such that $2(h+1)<\height{\B}$. We construct $\graypls{2}{h}{i}^{\height{\B}\times 2^{h}}$ as
\begin{equation*}
    \graypls{2}{h}{i}^{\intercal}=\begin{bmatrix}
        \mb{0}^{2^{h}\times i}&
        \bm{1}^{2^{h}\times 1}&
        \gray{2}{h}^{\intercal}&
        \mb{0}^{(2^{h}\times\height{\B}-i-h-1)}
    \end{bmatrix}^{\intercal}.
\end{equation*}
Then, $\B$ is constructed as follows:
\begin{equation*}\label{constructionB}
    \B=\begin{bmatrix}
        \bm{0}^{\height{\B}\times1}& \graypls{2}{h}{\height{\B}-1}&
        \hdots&
        \graypls{2}{h}{0}
    \end{bmatrix}.
\end{equation*}
\begin{example}\label{exampleB}
    The matrix $\B$ is constructed using $\gray{2}{2}$ and for $\height{\B}=7$:
\begin{align*}
    \begin{bmatrix}
    0&0011  &0110   &       &       &       &       &1111   \\
    0&0110  &       &       &       &       &1111   &0011   \\
    0&      &       &       &       &1111   &0011   &0110   \\
    0&      &       &       &1111   &0011   &0110   &       \\
    0&      &       &1111   &0011   &0110   &       &       \\
    0&      &1111   &0011   &0110   &       &       &       \\
    0&1111  &0011   &0110   &       &       &       &
    \end{bmatrix}
    \end{align*}
\end{example}
\begin{itemize}
    \item Condition~\ref{condition1}: we demonstrate a $2$-step procedure for recovering the index of each column based on its content which establishes that all columns are different. The idea is first to use $\bm{1}^{1\times 2^{h}}$ to recover $i$ (the index of $\graypls{2}{h}{i}$) and then use the following $h$ bits from the Gray code to locate the exact column. To do so, in the first step, we need an additional constraint $\height{\B}\geq 2(h+1)$. In a nutshell, we can recover $i$ by looking at the first $1$ after a length-$h+1$ burst of $0$ in each column. Moreover, by this constraint, each column in $\B$ has more zeros than ones. Consequently, all $\B(*,i)$ and $\bar{\B}(*,i)$ must be distinct.   %Since $\graypls{2}{h}{i}$ is all zero beside $\begin{bmatrix}
    %\bm{1}^{1\times2^{h}}\\
    %\gray{2}{h}
    %\end{bmatrix}$ part and $\height{\B}>2(h+1)$, one can distinguish each column by the following 2 steps. First, locate the start position $i$ by checking the first $1$ after the longest burst of $0$ in the column. Second, use the following $h$ bits from the gray code to locate the exact column. Moreover, each column in $\B$ has more zero than one. Consequently, all $\B\lp i\rp$ and $\bar{\B}\lp i\rp$ must be different. 
    \item Condition~\ref{condition2}: this condition is easy to verify.
    \item Condition~\ref{condition3}: by the construction from (\ref{constructionB}), each row $\B\lp i\rp$ is a horizontal cyclic shift of
    \begin{equation*}
        \bm{v}:= \lb \bm{1}^{2^{h}},\gray{2}{h}(0,*),\hdots,\gray{2}{h}(h-1,*)\rb
    \end{equation*}
    with an additional $0$ appended at the left. In Example~\ref{exampleB}, we have
    $\bm{v}=[1111 0011  0110]$.
    By this construction, the number of $-1$s in each row of $\shiftleft{\B}-\B$ equals the number of runs of $1$s in $\B(i,*)$, which equals the sum of the number of runs of $1$s in $\bm{1}^{1\times 2^{h}}$ and each row of the Gray code $\gray{2}{h}(i,*)$. By Fact~\ref{gray_num_burst}, the total number of runs of $1$s in each $\B(i,*)$ equals \vspace{-2mm}
    \begin{equation*}
        \sum_{i=1}^{h-1}2^{i-1} +2=2^{h-1}+1.\vspace{-2mm}
    \end{equation*}
\end{itemize}
Finally, we set $\thresholds{s}=2^{h-1}+2$. Then, $\frac{\ell}{\height{\B}}=\frac{\height{\B}2^{h}+1}{\height{\B}}>2^{h}= 2\thresholds{s}-4$ with the minor restriction that $\height{\B}>2(h+1)$.
 
\subsection{The Saturation SQGT Model for \Burstbnd}\label{Saturated_SQGT_bounded}
For the bounded-length burst problem \Burstbnd, one needs to recover both $\head{b}$ and $\tail{b}$ in order recover the burst. We describe next an order-optimal scheme for \Burstbnd restricted to the saturation SQGT model. First, for $\threshold{s}=s\geq l$, we describe an optimal scheme termed the \emph{Integer code}. Then, for $\threshold{s}=s<l$, by adapting the bursty GT scheme from~\cite{colbourn1999}, we arrive at an order-optimal scheme (within a constant factor of $4$).  
\begin{theorem}\label{Integer_code}
    For \Burstbnd and the saturation SQGT model for which $\thresholds=\lp0,\hdots,s\rp$ and $s\geq\ell$, there exists and order-optimal scheme $\mb{N}$ that solves \Burstbnd using $\lceil\log_{2}\lp n\rp\rceil + 1$ measurements.      
\end{theorem}
\begin{IEEEproof} The matrix $\mb{N}$ is a vertical concatenation of a $\lceil\log_{2}(n)\rceil\times n$ Index matrix and an $\bm{1}^{1\times n}$. Note that the $i$th column of the Index matrix is the binary representation of $i$.
    \begin{example}
For $n$=8, we have 
        \begin{align*}
        \mb{N}=\begin{bmatrix}
        0 & 1 & 0 & 1 & 0 & 1 & 0 & 1\\
        0 & 0 & 1 & 1 & 0 & 0 & 1 & 1\\
        0 & 0 & 0 & 0 & 1 & 1 & 1 & 1\\
        1 & 1 & 1 & 1 & 1 & 1 & 1 & 1
        \end{bmatrix}.
        \end{align*}
    \end{example}
    Since $\ell\leq s$, $\thresholds\lp\mb{N}\burst{b}\rp=\mb{N}\burst{b}$. We then treat the outcome vector as a binary representation of an integer $k$ equal to
    \begin{align*}
        k:&=\sum_{i=0}^{\lceil\log_{2}(n)\rceil}2^{i}\lp\mb{N}\burst{b}\rp(i,0)= \sum_{i=0}^{\lceil\log_{2}(n)\rceil}2^{i}\sum_{j=\head{b}}^{\tail{b}}\mb{N}(i,j)\\
        &= \sum_{j=\head{b}}^{\tail{b}}\sum_{i=0}^{\lceil\log_{2}(n)\rceil}2^{i}\mb{N}(i,j)= \sum_{j=\head{b}}^{\tail{b}}j= \frac{\head{b}+\tail{b}}{2}\length{b}.
    \end{align*}
We can deduce $\length{b}$ from the outcome corresponding to $\bm{1}^{1\times n}$. Hence, $\mb{N}$ recover the burst $\burst{b}$ using $\lceil\log_{2}(n)\rceil+1$ measurements.       
\end{IEEEproof}

\begin{theorem}\label{changes_colbourn_code}
    For \Burstbnd under the saturation SQGT model with $\thresholds=\lp0,\hdots,s\rp$ and $s<\ell$, there exists an order-optimal 
    scheme (within a constant factor $4$) $\mb{C}$ that solves \Burstbnd using $\leq\frac{2\ell}{s}+2\log_{2}(n)+3$ measurements.      
\end{theorem}
\begin{IEEEproof}
   The matrix $\mb{C}$ is as follows:
    \begin{equation*}
        \mb{C}=\begin{bmatrix}
            \mb{C}_{1}^{\intercal}&\mb{C}_{2}^{\intercal}&\mb{C}_{3}^{\intercal}
        \end{bmatrix}^{\intercal},
    \end{equation*}
where $\mb{C}_{1}$ is the Phase 1 matrix from Theorem 3.2 of~\cite{colbourn1999}. By an argument described in~\cite{colbourn1999}, 
$\mb{C}_{1}$ can distinguish all bursts $\burst{b}$ within distance $\geq 2(\ell-2)$. Next, we use $\mb{C}_{2}:=\lb\mb{I}_{\lceil\frac{2\ell}{s}\rceil}\otimes \bm{1}^{s}\rb^{\infty}$ truncated to $n$ columns from right. The outcome vector corresponding to $\mb{C}_{2}$ is a single run (in a circular sense) of non-zero of the form $o_{h},s,\hdots,s,o_{t}$, where $o_{h},o_{t}\in\{1,\hdots,s\}$. Then,
    \begin{enumerate}
        \item For $\length{b}>s$: since $\frac{\length{b}}{s}>1$ and $\lceil\frac{2\ell}{s}\rceil\geq\lceil\frac{2\length{b}}{s}\rceil\geq\lceil\frac{\length{b}}{s}\rceil+1$, $h\neq t$. Therefore, one can use $o_{h},o_{t}$ to recover $\head{b}\bmod 2\ell, \tail{b}\bmod 2\ell$. Consequently, $\mb{C}_{2}$ can distinguish all bursts at distance $<2\ell$ such that $\length{b}>s$.
        \item For $\length{b}\leq s$: from the outcome of $\mb{C}_{2}$, one can recover $\length{b}$. If $\length{b}\leq s$, then by Theorem~\ref{Integer_code}, $\mb{C}_{3}:=\mb{N}$ can distinguish all bursts with $\length{b}\leq s$.
    \end{enumerate}    
Therefore $\mb{C}$ can be used to solve \Burstbnd using $\height{\mb{C}_{1}}+\height{\mb{C}_{2}}+\height{\mb{C}_{3}}\leq \log_{2}(n)+\lceil\frac{2\ell}{s}\rceil+\lceil\log_{2}\lp n\rp\rceil + 1\leq\frac{2\ell}{s}+2\log_{2}(n)+3$ measurements, which is at most $4$ times the lower bound of Theorem~\ref{LB_variable}.    
\end{IEEEproof}

%\section{Algorithms}\label{sec:algo}
%\input{\srcpath sec_results.tex}
%\input{\srcpath sec_algo.tex}

%\section{Concluding Remarks}\label{sec:conclusion}
%\input{\srcpath sec_conclusion.tex}

%\section*{Acknowledgment}
%\input{\srcpath acknowledgment.tex}

\bibliographystyle{IEEEtran}
\bibliography{Ref.bib}

\begin{thebibliography}{10}
\providecommand{\url}[1]{#1}
\csname url@samestyle\endcsname
\providecommand{\newblock}{\relax}
\providecommand{\bibinfo}[2]{#2}
\providecommand{\BIBentrySTDinterwordspacing}{\spaceskip=0pt\relax}
\providecommand{\BIBentryALTinterwordstretchfactor}{4}
\providecommand{\BIBentryALTinterwordspacing}{\spaceskip=\fontdimen2\font plus
\BIBentryALTinterwordstretchfactor\fontdimen3\font minus
  \fontdimen4\font\relax}
\providecommand{\BIBforeignlanguage}[2]{{%
\expandafter\ifx\csname l@#1\endcsname\relax
\typeout{** WARNING: IEEEtran.bst: No hyphenation pattern has been}%
\typeout{** loaded for the language `#1'. Using the pattern for}%
\typeout{** the default language instead.}%
\else
\language=\csname l@#1\endcsname
\fi
#2}}
\providecommand{\BIBdecl}{\relax}
\BIBdecl

\bibitem{colbourn1999}
C.~J. Colbourn, ``Group testing for consecutive positives,'' \emph{Annals of
  Combinatorics}, vol.~3, no.~1, pp. 37--41, 1999.

\bibitem{emad2014semiquantitative}
A.~Emad and O.~Milenkovic, ``Semiquantitative group testing,'' \emph{IEEE
  Transactions on Information Theory}, vol.~60, no.~8, pp. 4614--4636, 2014.

\bibitem{dorfman}
R.~Dorfman, ``The detection of defective members of large populations,''
  \emph{The Annals of mathematical statistics}, vol.~14, no.~4, pp. 436--440,
  1943.

\bibitem{GT-ITperspective}
M.~Aldridge, O.~Johnson, J.~Scarlett \emph{et~al.}, ``Group testing: an
  information theory perspective,'' \emph{Foundations and
  Trends{\textregistered} in Communications and Information Theory}, vol.~15,
  no. 3-4, pp. 196--392, 2019.

\bibitem{combinatorial2000}
D.~Du, F.~K. Hwang, and F.~Hwang, \emph{Combinatorial group testing and its
  applications}.\hskip 1em plus 0.5em minus 0.4em\relax World Scientific, 2000,
  vol.~12.

\bibitem{lin2013synthetic}
Y.-L. Lin, C.~Ward, and S.~Skiena, ``Synthetic sequence design for signal
  location search,'' \emph{Algorithmica}, vol.~67, pp. 368--383, 2013.

\bibitem{muller2004consecutive}
M.~M{\"u}ller and M.~Jimbo, ``Consecutive positive detectable matrices and
  group testing for consecutive positives,'' \emph{Discrete mathematics}, vol.
  279, no. 1-3, pp. 369--381, 2004.

\bibitem{2021improved}
T.~V. Bui, M.~Cheraghchi, and T.~D. Nguyen, ``Improved algorithms for
  non-adaptive group testing with consecutive positives,'' in \emph{2021 IEEE
  International Symposium on Information Theory (ISIT)}.\hskip 1em plus 0.5em
  minus 0.4em\relax IEEE, 2021, pp. 1961--1966.

\bibitem{juan2008adaptive}
J.~S.-T. Juan and G.~J. Chang, ``Adaptive group testing for consecutive
  positives,'' \emph{Discrete mathematics}, vol. 308, no.~7, pp. 1124--1129,
  2008.

\bibitem{mutipleburst}
T.~V. Bui, Y.~M. Chee, J.~Scarlett \emph{et~al.}, ``Group testing with blocks
  of positives,'' in \emph{2022 IEEE International Symposium on Information
  Theory (ISIT)}.\hskip 1em plus 0.5em minus 0.4em\relax IEEE, 2022, pp.
  1082--1087.

\bibitem{emad2016code}
A.~Emad and O.~Milenkovic, ``Code construction and decoding algorithms for
  semi-quantitative group testing with nonuniform thresholds,'' \emph{IEEE
  Transactions on Information Theory}, vol.~62, no.~4, pp. 1674--1687, 2016.

\bibitem{lin2021positively}
Y.-J. Lin, C.-H. Yu, T.-H. Liu, C.-S. Chang, and W.-T. Chen, ``Positively
  correlated samples save pooled testing costs,'' \emph{IEEE Transactions on
  Network Science and Engineering}, vol.~8, no.~3, pp. 2170--2182, 2021.

\bibitem{AC-DC}
R.~Gabrys, S.~Pattabiraman, V.~Rana, J.~Ribeiro, M.~Cheraghchi, V.~Guruswami,
  and O.~Milenkovic, ``Ac-dc: Amplification curve diagnostics for covid-19
  group testing,'' \emph{arXiv preprint arXiv:2011.05223}, 2020.

\bibitem{cheraghchi2021semiquantitative}
M.~Cheraghchi, R.~Gabrys, and O.~Milenkovic, ``Semiquantitative group testing
  in at most two rounds,'' in \emph{2021 IEEE International Symposium on
  Information Theory (ISIT)}.\hskip 1em plus 0.5em minus 0.4em\relax IEEE,
  2021, pp. 1973--1978.

\bibitem{dyachkov1983survey}
A.~G. Dyachkov and V.~V. Rykov, ``A survey of superimposed code theory,''
  \emph{Problems of Control and Information Theory}, vol.~12, no.~4, pp. 1--13,
  1983.

\end{thebibliography}

%\newpage
%\onecolumn

%\appendix
%\input{\srcpath sec_proofs.tex}

%%%%%%
%% To balance the columns at the last page of the paper use this
%% command:
%%
%\enlargethispage{-1.2cm} 
%%
%% If the balancing should occur in the middle of the references, use
%% the following trigger:
%%
%\IEEEtriggeratref{3}
%%
%% which triggers a \newpage (i.e., new column) just before the given
%% reference number. Note that you need to adapt this if you modify
%% the paper.  The "triggered" command can be changed if desired:
%%
%\IEEEtriggercmd{\enlargethispage{-20cm}}
%%
%%%%%%

%%%%%%
%% References:
%% We recommend the usage of BibTeX:
%%

\end{document}